\documentclass[%
 aip,
 amsmath,amssymb,
 reprint,%
]{revtex4-1}

\usepackage{graphicx}
\usepackage{dcolumn}
\usepackage{bm}

\usepackage[utf8]{inputenc}
\usepackage[T1]{fontenc}
\usepackage{mathptmx}
\usepackage{etoolbox}
\usepackage{gensymb}
\usepackage {todonotes}

\makeatletter
\def\@email#1#2{%
 \endgroup
 \patchcmd{\titleblock@produce}
  {\frontmatter@RRAPformat}
  {\frontmatter@RRAPformat{\produce@RRAP{*#1\href{mailto:#2}{#2}}}\frontmatter@RRAPformat}
  {}{}
}%
\makeatother
\begin{document}
\preprint{AIP/123-QED}

\title{A structural analysis of ordered C\lowercase{s}$_{3}$S\lowercase{b} films grown on single crystal graphene and silicon carbide substrates}

\author{C. A. Pennington\textsuperscript{1}, M. Gaowei\textsuperscript{2}, E. M. Echeverria\textsuperscript{1}, K. Evans-Lutterodt\textsuperscript{2},\\ A. Galdi\textsuperscript{3}, T. Juffmann\textsuperscript{4,}\textsuperscript{5}, S. Karkare\textsuperscript{6}, J. Maxson\textsuperscript{1}, S.J. van der Molen\textsuperscript{7,10},\\ P. Saha\textsuperscript{2,6}, J. Smedley\textsuperscript{8}, W.G. Stam\textsuperscript{7,10}, R.M. Tromp\textsuperscript{9} \\
		\textsuperscript{1}Cornell Laboratory for Accelerator-Based Sciences and Education, Ithaca, NY 14850, USA \\
            \textsuperscript{2}Brookhaven National Laboratory, Upton, NY 11973-5000, USA \\
            \textsuperscript{3}University of Salerno, Department of Industrial Engineering, 84084 Fisciano (SA), Italy \\
            \textsuperscript{4}University of Vienna, Faculty of Physics, VCQ, A-1090 Vienna, Austria \\
            \textsuperscript{5}University of Vienna, Max Perutz Laboratories,\\ Department of Structural and Computational Biology, A-1030 Vienna, Austria\\
            \textsuperscript{6}Arizona State University, Tempe, AZ 85287, USA \\
            \textsuperscript{7}Leiden Institute of Physics, Niels Bohrweg 2, Leiden, The Netherlands\\
            \textsuperscript{8}SLAC National Accelerator Laboratory, Menlo Park, CA 94025, USA \\
            \textsuperscript{9}IBM T.J.Watson Research Center, Yorktown Heights, NY 10598, USA\\
            \textsuperscript{10}Leiden University, 2311 EZ Leiden, The Netherlands
}

\begin{abstract}
    Alkali antimonides are well established as high efficiency, low intrinsic emittance photocathodes for accelerators and photon detectors. However, conventionally grown alkali antimonide films are polycrystalline with surface disorder and roughness that can limit achievable beam brightness. Ordering the crystalline structure of alkali antimonides has the potential to deliver higher brightness electron beams by reducing surface disorder and enabling the engineering of material properties at the level of atomic layers. In this report, we demonstrate the growth of ordered Cs$_{3}$Sb films on single crystal substrates 3C-SiC and graphene-coated 4H-SiC using pulsed laser deposition and conventional thermal evaporation growth techniques. The crystalline structures of the Cs$_{3}$Sb films were examined using reflection high energy electron diffraction (RHEED) and X-ray diffraction (XRD) diagnostics, while film thickness and roughness estimates were made using x-ray reflectivity (XRR). With these tools, we observed ordered domains in less than 10 nm thick films with quantum efficiencies greater than one percent at 530 nm. Moreover, we identify structural features such as Laue oscillations indicative of highly ordered films. We found that Cs$_{3}$Sb films grew with flat, fiber-textured surfaces on 3C-SiC and with multiple ordered domains and sub-nanometer surface roughness on graphene-coated 4H-SiC under our growth conditions. We identify the crystallographic orientations of Cs$_{3}$Sb grown on graphene-coated 4H-SiC substrates and discuss the significance of examining the crystal structure of these films for growing epitaxial heterostructures in future experiments.
\end{abstract}
\maketitle

\section{Introduction}
Many modern accelerator applications employ high charge and ultra-short pulsed electron beams that are created by photoemission. These applications range from x-ray free electrons lasers (XFELs) to ultrafast electron diffraction and microscopy (UED/M) systems, which are both probes of matter at sub-nanometer spatial scales and femtosecond time scales\cite{MUSUMECI2018209}. The brightness of the electron beam is an important figure of merit that is used to determine flux and space-time resolution in these scattering experiments \cite{lobastov_four-dimensional_2005,Emma2010,Li_diffraction}.

In the case of XFELs, UED, and all other linear accelerators,  the maximum brightness achievable throughout the accelerator system can be no greater than at the source\cite{Bazarov_brightness, Pierce_intrinsic}. Desired properties of high brightness photocathode are those with high photon-to-electron conversion efficiency (quantum efficiency, QE), and those capable of emitting electron beams with small momentum spread in all directions. The latter property requires atomically smooth, ordered surfaces to minimize physical and chemical roughness that can lead to the creation of excess photoemission momentum spread\cite{Jfeng2023,Galdi_reduction,Cultrera_emittance,Gev_effects,dowell_quantum_2009,karkare_effects_2011}.
Recent studies identified 3C-SiC as a promising substrate for growing ordered cesium antimonide (Cs$_{3}$Sb) films \cite{galdi_understanding_2021, LaVia_3C}. Molecular beam epitaxy (MBE) with reflection high energy electron diffraction (RHEED) assisted deposition has successfully produced single crystal Cs$_{3}$Sb on 3C-SiC \cite{Parzyck_Cs3Sb}. These single crystal films demonstrate significantly higher quantum efficiency (QE), exceeding one percent at 530 nm for a 2 nm thick film, compared to their polycrystalline counterparts \cite{Parzyck_Cs3Sb}. Additionally, atomically flat alkali antimonide films, produced using sputtering techniques, have been grown at greater thicknesses than single crystal films \cite{Gaowei_sputter}. Other works have investigated correlations among crystallinity and photoemission performance of Cs$_{3}$Sb \cite{dowman_scanning_1975}.

In this work, we expand on previous results by using x-rays and RHEED to investigate the crystal structure of Cs$_{3}$Sb films grown on lattice-matched substrates to further understand correlations between film crystallinity,  quantum efficiency, and roughness. We use 3C-SiC as a lattice-matched substrate and also study a relatively new substrate for growth of Cs$_{3}$Sb films:  graphene-coated 4H-SiC (denoted Gr/4H-SiC), which can be prepared as a flat, defect-free surface surface \cite{chang_remote_2023,shi_van_2012,Guido_isr}. 

The films were characterized using a variety of synchrotron radiation-based methods and conventional thin film growth diagnostics. X-ray diffraction (XRD) and RHEED characterize the bulk and surface atomic structure, respectively.The chemical composition of the film is monitored by x-ray fluorescence (XRF) while the film thickness and roughness is measured with x-ray reflectivity (XRR). Finally, QE is measured at multiple wavelengths from 300 nm to 700 nm. The colocation of these tools, and the ability to simultaneously use or rapidly switch between them, is an important strength of our experimental configuration.

Understanding the crystal structure and conditions for epitaxial growth of alkali-antimonides is critical for developing high brightness photocathodes. Beyond the reduction of surface roughness described above, epitaxial growth opens the door to materials engineering on the scale of atomic layers by straining the crystal lattice or by heterostructuring. Exceptionally high spin-polarization was achieved from strained GaAs photocathodes\cite{Maruyama_strainedGaAs,pierce_photoemission_1976}. Spin-polarization has also been demonstrated in a study with alkali antimonides, which found that polarization can improve to greater than fifty percent through heterostructuring\cite{rusetsky_new}. QE enhancement via heterostructuring has also been demonstrated with GaAs and GaN photocathodes, but has yet to be demonstrated with the alkali antimonides\cite{Liu_GaAs,Marini2018PolarizationEN}.

\section{Experiment and Methods} 

\begin{figure}
    \centering
    \includegraphics[width=1.00\columnwidth]{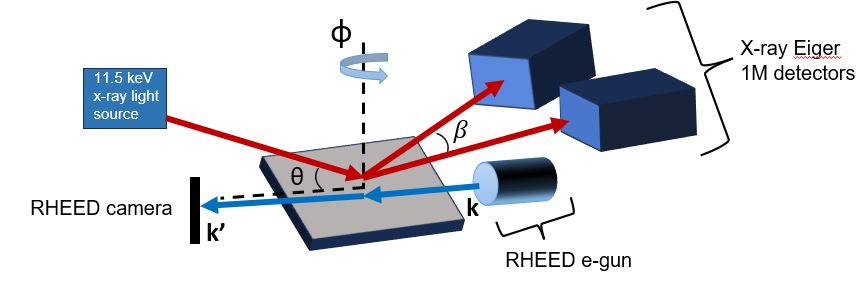}
    \caption{Schematic of x-ray and RHEED diagnostics inside the photocathode growth chamber. The 11.5 keV x-rays (red) from NSLS-II scatter off the film at the characteristic angle $\theta $. X-ray cameras are positioned to capture high and low angle diffraction. Electrons from the 15 kV RHEED system interact with the film surface at a grazing incidence angle of less than 2 degrees. The film sample can be rotated about the angle $\phi$ in the chamber on a motorized stage.}
    \label{diagnostics}
\end{figure}

We grew thin films of Cs$_{3}$Sb in a dedicated deposition chamber at the National Synchrotron Light Source II (NSLS-II) Beamline 4-ID ISR of Brookhaven National Laboratory. The growth chamber was mounted at the end of the x-ray beamline on a motorized rotation stage. We fix the x-ray energy at 11.5 keV (1.08 \AA). X-rays enter the vacuum chamber via beryllium windows, interact with the film sample, and exit the chamber. Two Eiger 1M cameras shown in FIG.~\ref{diagnostics} were positioned behind the beryllium windows; the first camera captures grazing angle 
scattering for the XRR and coplanar diffraction and the second camera was positioned at a closer distance to the chamber and 30 degrees away from the first camera (angle $\beta$ in FIG.~\ref{diagnostics}) to capture higher 
angle Bragg reflections. The chamber itself was mounted on a motorized stage, which enabled us to scan the 2$\theta$ angle for sampling diffraction modes. Chamber pressures were maintained at $1 \times 10^{-9}$ Torr during film deposition with a base pressure of $2 \times 10^{-10}$ Torr. 

The substrates were cleaned with isopropyl alcohol (IPA) and methanol sonication prior to growth, then placed on a holder in the center of the growth chamber that could rotate about the substrate normal by approximately 30 degrees (angle $\phi$ in FIG.~\ref{diagnostics}) . The substrates were then annealed between 30 minutes and 1 hour at 500-550$\degree$C.

The Cs$_{3}$Sb samples were grown on the 3C-SiC and Gr/4H-SiC substrates, as well as Gr/SiO$_{2}$ as a test substrate for a comparison to Gr/4H-SiC. The graphene layers on the 4H-SiC were epitaxially grown on 4H-SiC in the IBM LEEM-II instrument\cite{Tromp_graphene,de_Jong_stacking,Han_graphene,Hass_graphene}, while the graphene layer was transferred onto the SiO$_{2}$.

Two thin film growth techniques were used: pulsed laser deposition (PLD) and conventional thermal evaporation\cite{ashford_PLD}. The PLD method uses a pulsed excimer laser (248 nm) incident on a commercial Sb target to vaporize Sb onto the mounted substrate. An in-vacuum resistive heating cell was used to evaporate Sb in for thermal evaporation. A similar resistive heating cell was used to evaporate Cs into the chamber for both growth techniques. 

Two deposition methods were used during the growth on 3C-SiC substrates: co-deposition of Cs and Sb at a single temperature, and a dual temperature growth using a deposition temperature and an annealing temperature. Single-temperature co-deposition was the only technique used for the films grown on Gr/4H-SiC. During single-temperature co-deposition, the substrate was held at a constant temperature between 80-100$\degree$C while Cs and Sb were simultaneously deposited on the substrate. During the two-temperature growth, the substrate was held at temperatures between 40-60$\degree$C in the deposition step until approximately 2-3 monolayers of Cs$_{3}$Sb were deposited. The sample was then heated to 80-110$\degree$C in the annealing step for 15 minutes. The Cs shutter was left open to replace de-sorbed Cs, while the Sb flux was turned off in this step. This cycle completes a deposition stage. We completed 2-4 deposition stages per sample, and performed XRD, XRR, RHEED, and QE scans after each deposition stage. The XRD provided structural information in the bulk of the sample, RHEED provided structural information at the surface, XRR provided film thickness measurements and rms surface roughness estimates, and the QE diagnostic provided measurements of the number of electrons emitted from the film divided by the number of incident photons. Growth parameters such as temperatures and deposition rates for each sample will be discussed in the following sections. 

The x-ray diagnostics used in this experiment were similar to those described in \cite{Gaowei_sputter,ruiz_oses_bialkali,Xie_Synchrotron}. To measure QE, a laser-driven light source (LDLS) and monochromator were used to illuminate the photocathode films over a 300-700 nm wavelength range and photocurrent was measured by inserting a biased anode at 50-80 V approximately 5 mm from the cathode surface.

 \begin{figure*}
    \centering
    \includegraphics[width=2.00\columnwidth]{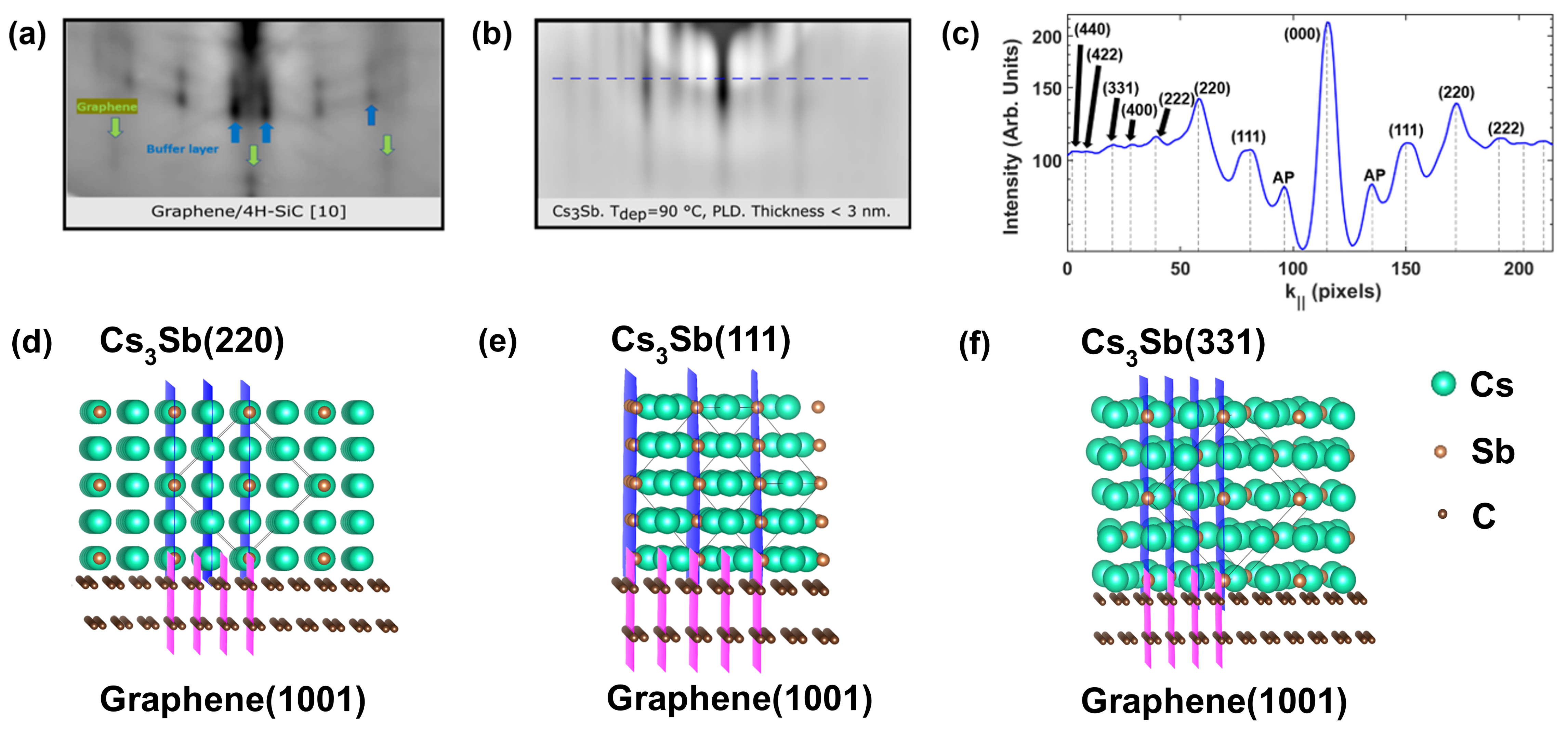}
    \caption{Growth and surface structure characterization of Cs$_{3}$Sb films grown on Gr/4H-SiC using PLD. (a) RHEED image of Gr/4H-SiC along the [10] direction. Reflection streaks from graphene (green arrows) are seen as well as satellite reflections from the carbon buffer layer\cite{goler_revealing_2013,cavallucci_intrinsic_2018} (blue arrows) that lies between the graphene and SiC layers are seen. (b) a RHEED pattern of the Cs$_{3}$Sb film. The position of the dashed line (blue) is used to generate the intensity profile in (c). Distances between reflection peaks were used to calculate the corresponding d-spacing and azimuth planes labeled above each peak. The first order peaks are labeled `AP' for another phase of the material, as it does not correspond to a possible d-spacing of Cs$_{3}$Sb and is further described in the text. We estimate the thickness of the film to be less than 3 nm with XRR. Schematics of probable orientations that could result in epitaxy between the film and substrate are shown for (d) Cs$_{3}$Sb(220) (e) Cs$_{3}$Sb(111), and (f) Cs$_{3}$Sb(331) with matching lattice-plane periodicity between Cs$_{3}$Sb (blue) graphene (purple) and with corresponding unit cells (black lines).}
    \label{RHEED_Gr/4H}
\end{figure*} 

\section{Results and Discussion}



We first look at the RHEED patterns from the Gr/4H-SiC substrate and Cs$_{3}$Sb film in FIG.~\ref{RHEED_Gr/4H}. Qualitative information about the surface structure can be interpreted from the RHEED pattern such as whether the surface is smooth, rough, or textured. The RHEED pattern also provides quantitative information, in particular the d-spacing between crystal faces which can be calculated from the distance between reflections. 

To identify the in-plane crystallographic orientations, we use RHEED analyses similar to Hafez and Szata to calculate the d-spacing of the film and characterize the surface\cite{Hafez_RHEED,Szata_RHEED}. We measure the distance between the streaks in the pattern, $l_{n}$, and determine the distance between the lattice vectors in reciprocal space, $k$, using the equation

\begin{equation}
    nk = \frac{k_{e}}{\sqrt{(\frac{L}{l_{n}})^{2}+1}}
        \label{krods}
\end{equation}

\noindent where $k_{e}$ is the electron wavenumber ($k_{e} = \frac{2 \pi}{h}\sqrt{2mE+\frac{E^{2}}{c^{2}}}$, $E$ is the electron kinetic energy, h is Planck's constant, and $L=27$ cm is the distance from the sample to the RHEED detector. The d-spacing can then be calculated as $d = \frac{2 \pi}{k}$, where $k$ is calculated from Equation \ref{krods}. We use the first diffraction order $n=1$ for all calculations of d-spacing. The RHEED beam energy is 15 keV. 

The RHEED pattern of the Gr/4H-SiC substrate in FIG.~\ref{RHEED_Gr/4H}(a) consists of reflections from the graphene surface layer (green arrows) as well as `satellite' peaks from the carbon buffer layer  \cite{goler_revealing_2013,cavallucci_intrinsic_2018} (blue arrows) that lies between the graphene layers and 4H-SiC surface. We measure the distance between the graphene peaks in the substrate image and find a d-spacing of 2.20 \textup{~\AA} which corresponds to the (1001) lattice plane (measured value: d = 2.19 \textup{~\AA})\cite{Sheha_graphene,zagorac_recent_2019} which we will denote the [10] direction. 

We use the same analysis using the RHEED image of the approximately 3 nm thick film grown with PLD shown in FIG.~\ref{RHEED_Gr/4H}(b), where an intensity profile is extracted along the dashed horizontal line and plotted in FIG.~\ref{RHEED_Gr/4H}(c). The d-spacing values were calculated using Equation \ref{krods} and corresponding Miller indices \textit{(hkl)} are identified for the interplanar spacing along the azimuth as in the case of the substrate. Several in-plane Cs$_{3}$Sb crystal orientations relative to the (1001) orientation of hexagonal graphene were observed. 

We also find a d-spacing which does not match an allowed d-spacing of FCC Cs$_{3}$Sb in FIG.~\ref{RHEED_Gr/4H}(c). This single reflection could appear from a domain of Cs$_{3}$Sb$_{7}$(111), where the calculated d-spacing of \textit{d} = 10.1 \textup{~\AA}\cite{SingleCrystal,hirschle_darstellung_2000} is approximately the same as the  measured d-spacing, \textit{d} = 10.2 \textup{~\AA}. Another likely explanation for this peak is surface reconstruction. 

A schematic of possible epitaxial orientations of Cs$_{3}$Sb observed in the experiment are shown in FIG.~\ref{RHEED_Gr/4H}(d),(e),(f). The separation between lattice planes of the film and substrate approximately correspond to integer multiples in each of the observed cases from FIG.~\ref{RHEED_Gr/4H}, confirming that multiple epitaxial orientations are likely for this film and substrate pair. We use the ideal F$_{m\bar{3}m}$ structure of Cs$_{3}$Sb for visualization purposes instead of the identified stable structure which has approximately the same lattice constant \cite{nangoi2022ab, Parzyck_Cs3Sb}. 
 
Qualitatively, the vertical streaks in the RHEED pattern for the film indicate that the surface is very smooth. This is confirmed with x-ray reflectivity (XRR) measurements for multiple thicknesses of the grown film from 3-20 nm\cite{Guido_isr}, demonstrating that while the films grew with multiple grain orientations on the substrate, the rms roughness remains less than 1 nm for film thicknesses up to 30 nm. 

\begin{figure}
    \centering
    \includegraphics[width=1.00\columnwidth]{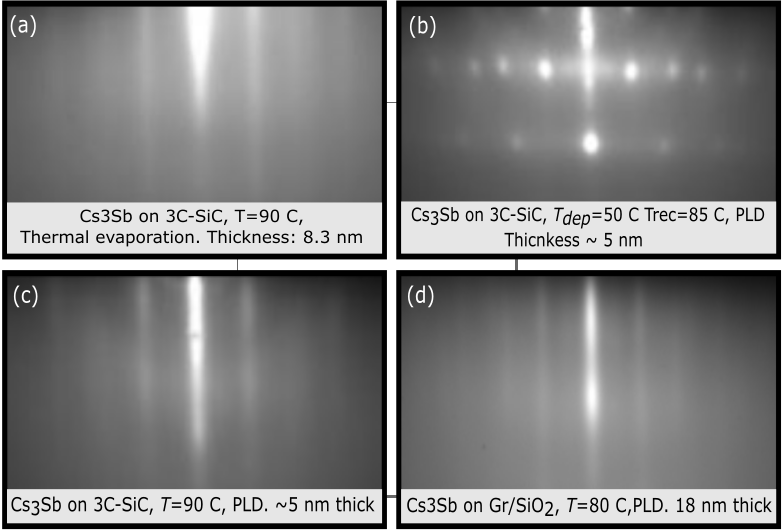}
    \caption{RHEED images of Cs$_{3}$Sb on 3C-SiC(100) and SiO$_{2}$ substrates.(a) A Cs$_{3}$Sb film with a fiber textured surface grows on 3C-SiC, (b) decreasing the deposition temperature and using a two temperature deposition method results in the growth of an ordered film with the formation of 3D islands (roughness). (c) Smooth, fiber textured films are grown on 3C-SiC using PLD, and (d) a fiber textured film is grown on Gr/SiO$_{2}$ with PLD.} 
    \label{RHEED_fibertexture}
\end{figure}

In FIG.\ref{RHEED_fibertexture}, we present RHEED patterns of films grown on 3C-SiC using both PLD and thermal evaporation techniques. The streak patterns in FIG.\ref{RHEED_fibertexture}(a) indicate that a smooth surface was achieved under the specified growth parameters. Conversely, a RHEED pattern consisting of spots is observed in FIG.~\ref{RHEED_fibertexture}(b) from a film grown using a two-temperature growth method with a deposition temperature of 50$^\circ$C and an annealing temperature of 85$^\circ$C. This pattern of transmission spots indicates an ordered film was grown with three-dimensional 'islands' on the surface, and is visually distinct from a pattern corresponding to an extremely smooth, single crystal film, which would consist of small spots along a Laue circle. These results suggest we have yet to optimize the two-temperature profile to achieve the single crystal film quality demonstrated by Parzyck and Galdi\cite{Parzyck_Cs3Sb}. 

In FIG.~\ref{RHEED_fibertexture}(c), we show the RHEED pattern of a film grown on 3C-SiC using PLD. This RHEED pattern consists of smooth streaks and is qualitatively identical to the pattern in FIG.~\ref{RHEED_fibertexture}(a). The resulting films were fiber textured: the surface was ordered out-of-plane while the domain orientations were misaligned in-plane. This result, along with the results from the films grown on Gr/4H-SiC, indicates that the substrate has a larger influence on the resulting crystal structure of the film than the choice of PLD or thermal evaporation. 

Finally, we compare the ordered films grown on 3C-SiC to a film grown on a SiO$_{2}$ substrate with a bilayer of graphene on the surface, denoted as Gr/SiO$_{2}$, in FIG.~\ref{RHEED_fibertexture}(d). The RHEED pattern of the Cs$_{3}$Sb film grown on the Gr/SiO$_{2}$ substrate reveals that the film is fiber textured, and qualitatively similar to the films grown on 3C-SiC. However, when comparing the quality of the films grown on Gr/SiO$_{2}$ to the films grown on Gr/4H-SiC in FIG.~\ref{RHEED_Gr/4H}, which had an ordered in-plane structure on the surface, the difference is likely due to Gr/4H-SiC having a smoother surface than Gr/SiO$_{2}$, where the graphene layer was transferred to SiO$_{2}$. Similar structural differences between Cs$_{2}$Te films grown on Gr/4H-SiC and Gr/SiO$_{2}$ have been observed in other experiments\cite{Kali_CsTe}. 


\begin{figure}
    \centering
    \includegraphics[width=1.00\columnwidth]{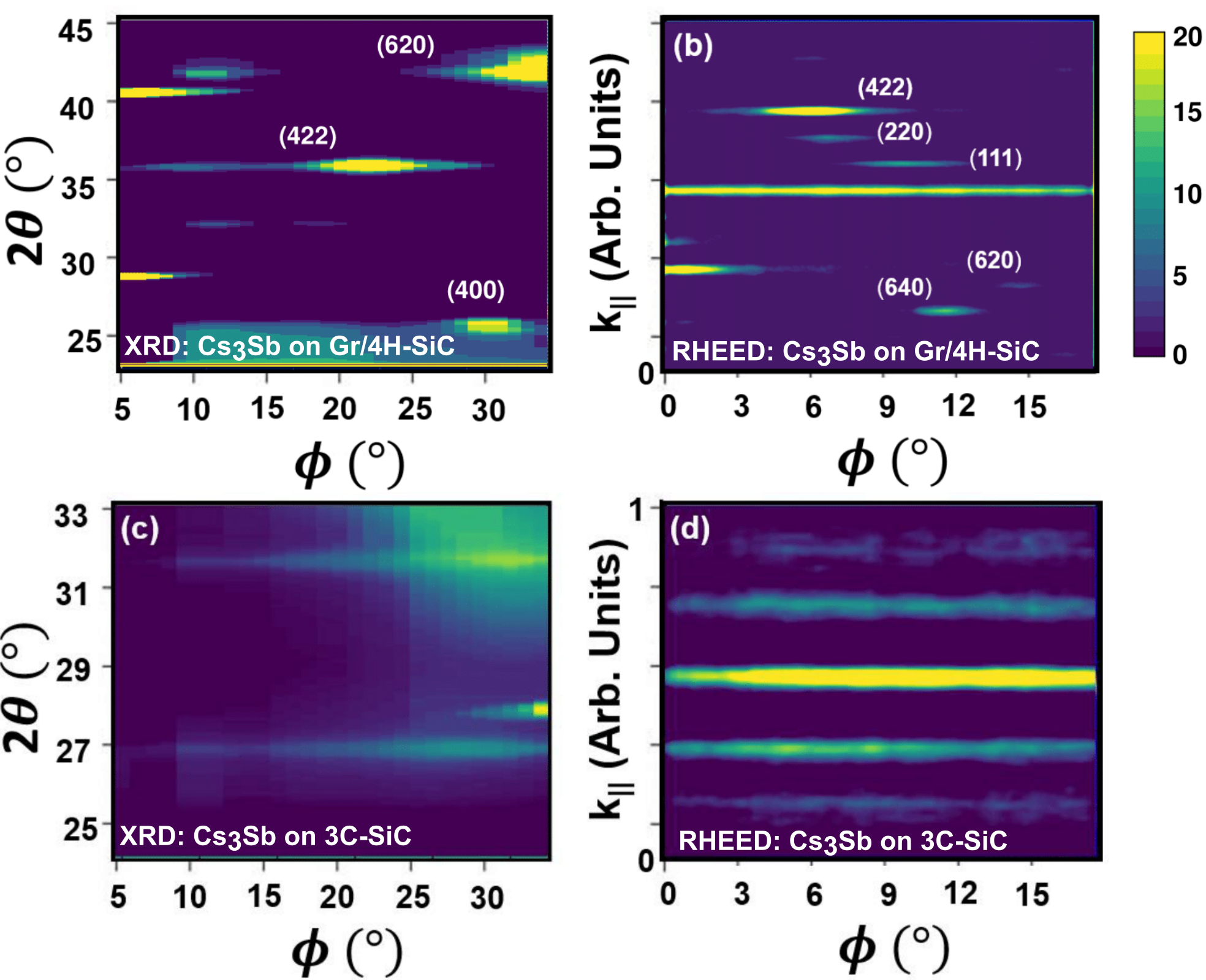}
    \caption{(a) the XRD diffraction peaks as a function of angle $\phi$ as the Cs$_{3}$Sb film on Gr/4H-SiC is rotated about the azimuth. Rotational dependence of the XRD peaks implies the growth of ordered domains, which are identified by the Miller indices (400), (422), and (620). (b) the azimuth scan of the film with RHEED identifies the same surface azimuths as in (a) in addition to others.(c) XRD captures small modulations in the film on 3C-SiC and (d) RHEED find no modulations on the surface of the films grown on 3C-SiC indicating smooth and fiber textured surfaces. }
    \label{RHEED_Gr/4H_throck}
\end{figure}

To determine the crystallinity of the films, we examine the azimuthal dependence of the RHEED and XRD patterns of the deposited Cs$_{3}$Sb film relative to the Gr/4H-SiC and 3C-SiC substrates, as shown in FIG.~\ref{RHEED_Gr/4H_throck}. Modulations in the RHEED and XRD azimuth scans observed in FIG.~\ref{RHEED_Gr/4H_throck} imply in-plane ordering in the thin films. The maximum angular range in the azimuth scans was approximately 30$^\circ$, but even in this limited range we were able to capture sharp intensity modulations from the RHEED and XRD scans for films grown on Gr/4H-SiC. The labeled intensity modulations in the the XRD azimuth scan shown in FIG.~\ref{RHEED_Gr/4H_throck}(a) correspond to three different grain orientations of a 20 nm thick film on Gr/4H-SiC grown with PLD, which is from a subsequent growth on the sample sample shown in FIG.~\ref{RHEED_Gr/4H}. Signatures of these same grain orientations, along with others, are observed in the RHEED scan of this film in FIG.~\ref{RHEED_Gr/4H_throck}(b), which are identical to the sample in FIG.~\ref{RHEED_Gr/4H}.

The azimuth scans of an 8.3 nm thick film grown by thermal evaporation on 3C-SiC are shown in FIG.~\ref{RHEED_Gr/4H_throck}(c),(d), where the XRD scan (FIG.~\ref{RHEED_Gr/4H_throck}(c)) shows azimuthal modulations for film peaks and the RHEED scan in FIG.~\ref{RHEED_Gr/4H_throck}(d) shows an absence of modulation. The absence of modulations with streaks in the RHEED image indicate that the surface is fiber textured. This is an intriguing result, since the x-rays detect intensity modulations about the angle $\phi$ while the RHEED does not. An interpretation of this result is that the 8.3 nm thick film was ordered in-plane below the surface, at an earlier stage in the growth, and became less ordered as the thickness increased. Another possibility is that the XRD was more sensitive to partial ordering than the RHEED in this case.  

\begin{figure}
    \centering \includegraphics[width=1.00\columnwidth]{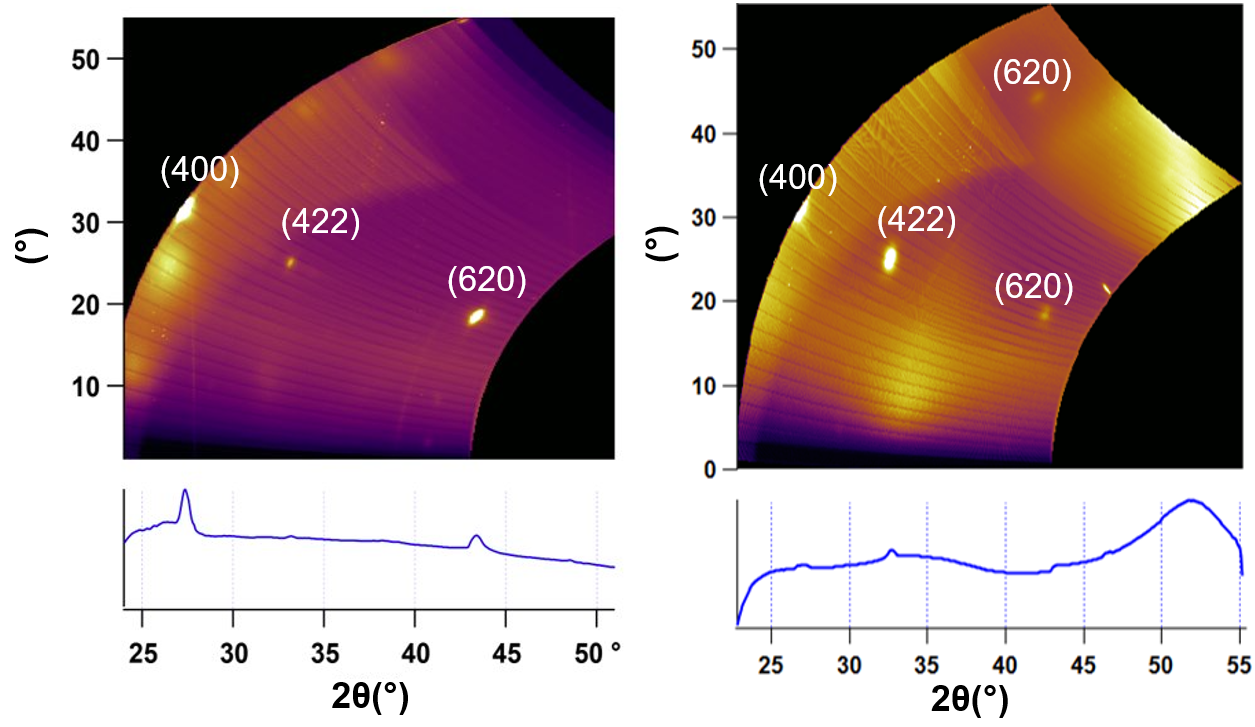}
    \caption{2$\theta$ scans of Cs$_{3}$Sb grown on (a) Gr/4H-SiC via PLD, film
thickness: 10 nm, and (b) 3C-SiC, film thickness: 8.3 nm, grown via
thermal evaporation. The blue curves are the integrated intensities
along the tilt axis. Some Bragg peaks are seen in identical locations
for the films grown on different substrates. The 3C-SiC substrate
Bragg peaks are the bright spots at 2$\theta$ $\approx$ 34$^{\circ}$ and at 2$\theta$ $\approx$ 53$^{\circ}$.}
    \label{XRD_Gr_3C}
\end{figure}

We now examine the 2$\theta$ scans of the films grown on 3C-SiC and Gr/4H-SiC. The X-ray 2$\theta$ scan of the film on Gr/4H-SiC is depicted in FIG.~\ref{XRD_Gr_3C}(a), where the Miller indices \textit{(hkl)} are indicated in white. In this scan, we observe Bragg reflections corresponding to the (400), (422), and (620) planes of the film, relative to the (100) plane of the substrate. The vertical axis of FIG.~\ref{XRD_Gr_3C}(a) represents the angular range of the detector, while the horizontal axis shows the range of 2$\theta$ values. This figure was generated by stitching together two-dimensional images over a range of 2$\theta$ values to provide a comprehensive view of the diffraction pattern.


In FIG.~\ref{XRD_Gr_3C}(b) we see several Bragg peaks appearing in identical locations for the samples grown on both Gr/4H and 3C-SiC. This implies that the Cs$_{3}$Sb films grew with some of the same out-of-plane grain orientations on both cubic 3C-SiC surfaces and hexagonal Gr/4H-SiC surfaces. While the films did not grow as a single crystal, they also did not grow purely polycrystalline, but instead as multiple large crystallites with surface and out of plane ordering. The ordered domains in the film are large enough to produce tight Bragg peaks in the 2$\theta$ scans, but small compared to the millimeter probe size to see multiple orientations. 

\begin{figure}
    \centering    \includegraphics[width=1.0\columnwidth]{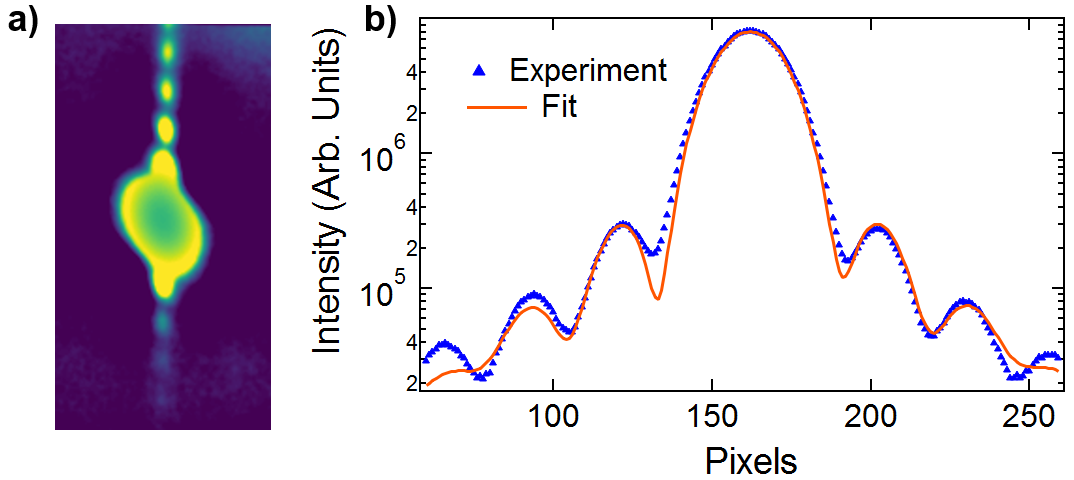}
    \caption{ (a) The (422) Bragg peak on camera 2 from a 10.5 nm thick film grown on Gr/4H-SiC. Laue oscillations appear around the film confirming the growth of flat, crystalline films. (b) A line profile of the oscillations about the peak in the vertical direction are plotted along with a fitted sinc$^{2}$($\theta$) function to show the behavior is of the correct functional form\cite{Miller_Laue}.}
    \label{XRD_fringes}
\end{figure}

\begin{figure*}
    \centering  \includegraphics[width=2.0\columnwidth]{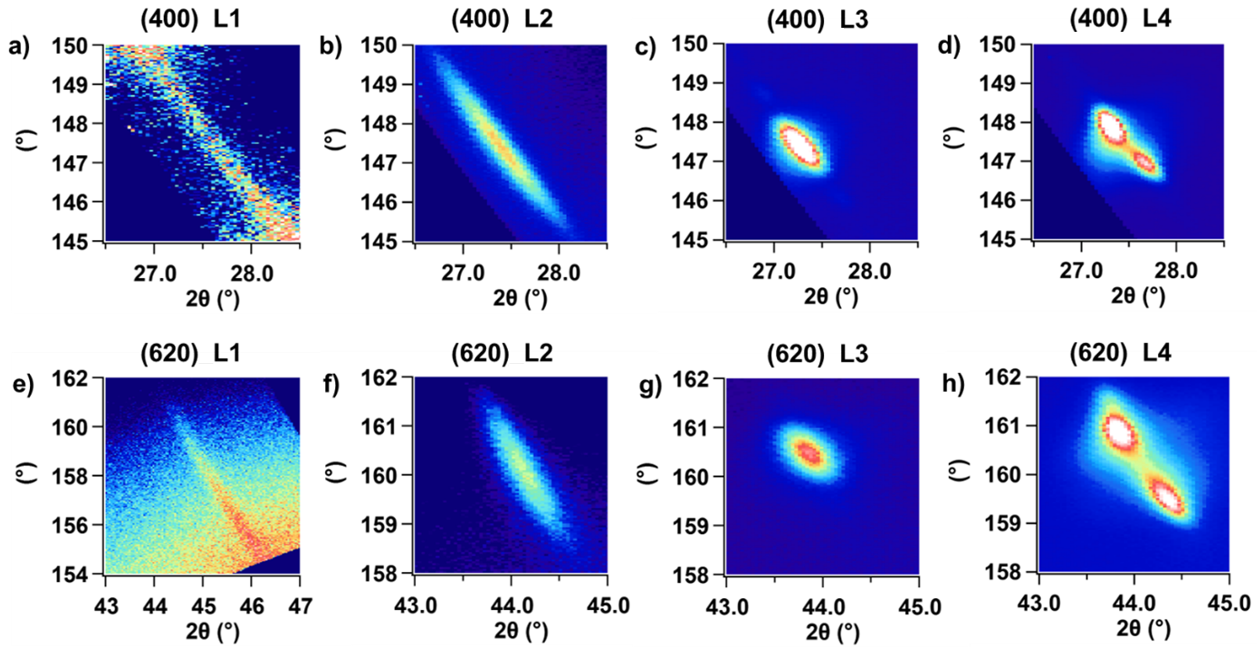}
    \caption{Diffraction peak (400) (a) through (d) and (620) (e) through (h) labeled by first to final layer, L1 to L4 respectively, of a Cs$_{3}$Sb film grown on Gr/4H-SiC using PLD. Film thicknesses are L1: 2 nm, L2: 4nm, L3: 10 nm and L4: 20 nm. The crystal structure becomes more defined as the layer thickness increases to 10 nm. Bragg peak splitting is seen at 20 nm thickness as tensile strain releases in the film.}
    \label{peak_split}
\end{figure*}

Another feature we observed from the Cs$_{3}$Sb films on Gr/4H-SiC is Laue oscillations around Bragg peaks shown in FIG.~\ref{XRD_fringes}. Laue oscillations result from incomplete destructive interference of the Bragg reflections when the domains of the probed film consist of the same number of unit cells, and are generally indicative of smooth, high quality crystalline films  \cite{Miller_Laue,Tetsuhiko_laue}. We fit a sinc$^{2}$($\theta)$ function to a vertical line cut of the oscillations from FIG.~\ref{XRD_fringes}(a) and shown in FIG.~\ref{XRD_fringes}(b) to show the feature is of the correct functional form, confirming the growth of flat, structurally ordered thin films. We estimate the thickness to be 10.5 nm from Equation \ref{laue_osc}\cite{Wainfan_laue}:

\begin{equation}
\Delta t \approx \frac{2 \pi}{\Delta q}
\label{laue_osc}
\end{equation}

where $\Delta$t is the thickness of the film and $\Delta$q is the distance from the Bragg peak to the furthest Laue fringe in inverse nanometers, which agrees with the 10.5 nm thickness measurements from XRR.

We now look at the evolution of the film structure in FIG.~\ref{peak_split}, with the XRD scans showing Bragg peak (400) and (620) at film thicknesses of 2 nm, 4nm, 10 nm and 20 nm, labeled L1-L4 respectively. The broadened shape in the diffraction peaks from L1 and L2 are most likely coming from misfit dislocations that occurred in the heteroepitaxial layers due to lattice mismatch and relaxation. In L3, both diffraction peaks are showing increased intensity and reduced peak width, indicating reduced stress stored in the crystal lattices and the crystalline domains in the film (in the probing diffraction plane). As the film thickness increases and the stress releases in the film, twist and tilt occurs in the crystalline domains, resulting in increased mosaicity which can be seen by multiple peaks in the diffraction pattern in FIG.~\ref{peak_split}(d),(h) \cite{Dolabella_strain,Graca_mosaicity}. 

Bragg peak splitting in L4 of FIG.~\ref{peak_split} occurred for films grown on 4H-SiC at thicknesses of approximately 20 nm, suggesting the formation of a new crystal grain in the film due to strain relaxation with possible defects. Surface defects along with loss of coordination with the substrate at greater thicknesses influenced the crystal grains to grow with less strain compared to the grains produced at lesser thickness. In both (400) and (620) cases, the newly formed Bragg peak in L4 appeared at a larger value of 2$\theta$ and smaller d-spacing from \textit{d} $\approx$ 2.31\textup{~\AA} in L3 to d $\approx$ 2.28 \textup{~\AA} for the new peak in L4 indicating the release of tensile strain with increasing film thickness.  

\begin{figure}
    \centering
    \includegraphics[width=1.00\columnwidth]{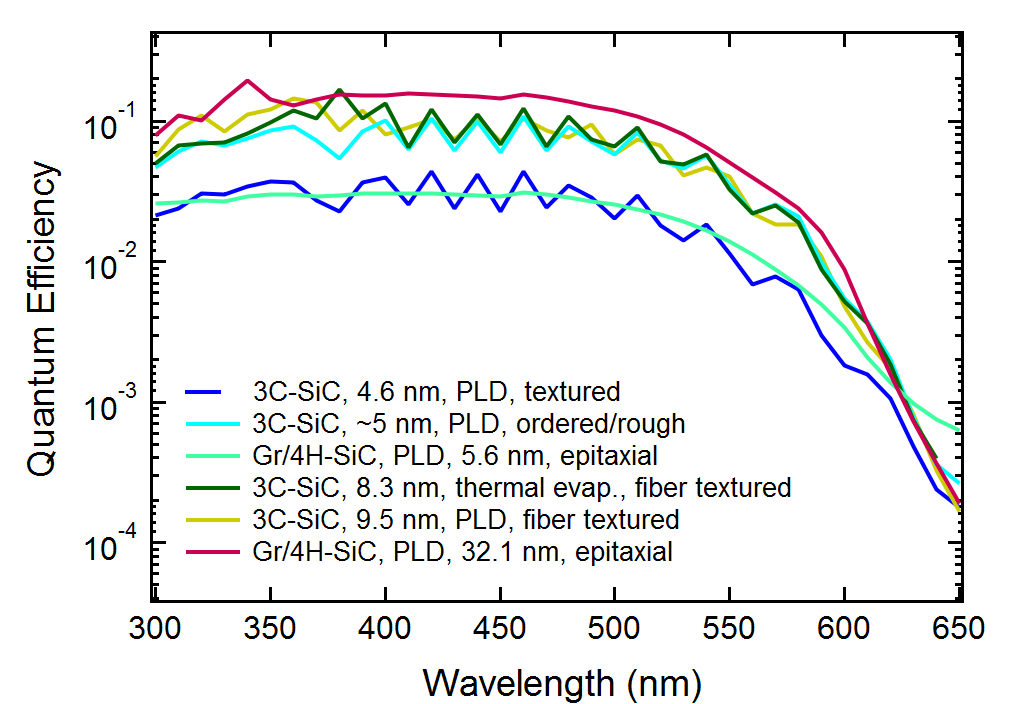}
    \caption{Spectral response of Cs$_{3}$Sb films grown on 3C-SiC and Gr/4H-SiC via PLD and thermal evaporation. Listed in the legend are the substrates, film thickness, growth technique, and film surface characterization from RHEED for each film. }
    \label{QE}
\end{figure}

In addition to investigating the surface and bulk structure of ordered Cs$_{3}$Sb films, we collected the spectral response over a range of ultraviolet and visible wavelengths as shown in FIG.~\ref{QE}. The measured QE was in excess of one percent at 530 nm for the thin ordered films grown on 3C-SiC.  

The modulations in the quantum efficiency seen in FIG.~\ref{QE} might initially appear to be noise in the measurement, but they actually arise from optical interference effects in the multilayer substrate as described by Alexander\cite{Alexander_enhanced}. The 3C-SiC substrate is composed of a silicon carbide layer and a silicon layer, where the interfaces of the multilayer reflect light back into the overlaying photocathode. This reflection leads to constructive or destructive interference with the incident light, producing the observed oscillations. We find that this cathode-substrate system improves the quantum efficiency by more than a factor of two at constructively interfering wavelengths compared to a cathode without QE enhancement from optical interference. The maximum quantum efficiency can be set to a desired wavelength in this optical system by tuning the thicknesses in the multilayer\cite{Pennington_QE}.

Looking across data sets, the film surface roughness did not have a measurable effect on the QE, and neither did the growth technique (PLD or thermal evaporation) or the crystal ordering of the film. QE dependence on film thickness is the only correlation that was observed in addition to optical interference\cite{Guido_isr}. 

An advantage of growing atomically-ordered films with low roughness compared to polycrystalline Cs$_{3}$Sb is the potential reduction in intrinsic emittance as described by Galdi\cite{Galdi_reduction}. The silicon carbide substrates facilitate the growth of thin, ordered Cs$_{3}$Sb films with QE in excess of one percent at 530 nm and several percent at 400 nm. A notable advantage of growing these films on lattice-matched substrates and utilizing the XRD and RHEED structural diagnostics is that ordered films of Cs$_{3}$Sb with high QE are consistently reproducible and can be grown within minutes after suitable growth parameters are identified.

\section{Summary and Conclusion}

In summary, highly ordered Cs\(_{3}\)Sb films were synthesized on graphene (Gr/4H-SiC) and silicon carbide (3C-SiC) substrates using pulsed laser deposition (PLD) and conventional thermal evaporation growth techniques. X-ray diffraction (XRD) and reflection high-energy electron diffraction (RHEED) were used to observe the crystal structure of the films. Ordered domains were observed with both in-plane and out-of-plane preferential ordering on the Gr/4H-SiC substrate for the first time, while smooth, fiber-textured films with out-of-plane ordering were observed on 3C-SiC. XRD and RHEED diagnostics were used to determined the surface and normal orientations of the Cs\(_{3}\)Sb domains relative to the (1001) orientation of the Gr/4H-SiC substrate.

For the films grown on 3C-SiC, XRD scans showed tight, well-defined Bragg peaks that had subtle azimuthal dependence, while RHEED images indicated out-of-plane ordering with no azimuthal dependence and a disordered in-plane structure. The tight Bragg peaks in the XRD 2\(\theta\) results, combined with the lack of long-range ordering from RHEED results, suggest that the film was more ordered in the early stages of growth and became disordered in-plane as the thickness increased.

This claim is further supported by the Bragg peak splitting observed for films grown on Gr/4H-SiC with thicknesses near 20 nm, indicating changing strain in the films. The formation of an additional Bragg peak at a slightly larger 2\(\theta\) angle indicates the formation of a domain with a larger d-spacing and thus relaxation of tensile strain. Greater strain at earlier stages of the growth then suggests that there was coordination between the substrate structure and film structure.

Laue oscillations around the Bragg peaks were seen for films grown on Gr/4H-SiC at thicknesses near 10 nanometers, indicating the growth of a well ordered film and high out-of-plane coherence. A calculation of the film thickness was performed using this feature, which agreed with thickness measurements from x-ray reflectivity. This agreement implies that the vertical, out-of-plane ordering of the film is coherent throughout the entirety of the 10 nanometer thickness. Combining this conclusion with the in-plane and out-of-plane ordering observed with RHEED and XRD, respectively, we conclude that these films grew as large crystallites with a few different orientations relative to the substrate, but with exceptionally high in-plane and out-of-plane coherence within each crystallite domain. 

In addition to the structural characterization, quantum efficiency (QE) of the crystalline films was measured across a range of UV and visible light, showing percent-level QE at 530 nm. An optical interference effect in the photocathode-substrate multilayer enhanced the QE by more than a factor of two at wavelengths coinciding with constructive interference from this feature. 


The significance of examining the crystal structure of alkali antimonides using XRD and RHEED is twofold: Firstly, it validates the ability to synthesize films with improved surface structure and reduced roughness compared to conventionally grown polycrystalline films using PLD and thermal evaporation. Secondly, it demonstrates that a comprehensive 3D structural characterization of ordered alkali antimonides can be performed, which is advantageous for future experiments aiming to grow heterostructures with these compounds, as it provides insights into crystal ordering at various stages of growth. This structural understanding can be effectively provided by XRD and RHEED diagnostics, as demonstrated in this study. These findings underscore the capability of XRD and RHEED diagnostics in enabling the consistent growth of smooth and ordered alkali antimonide films with high quantum efficiency within minutes, once suitable growth parameters are identified.

\section{Acknowledgements}

The authors thank Vitaly Pavlenko, Jean Jordan-Sweet, John Walsh, and Randall Headrick for their technical support at the beamline and helpful discussions during the experiments.
This research used resources 4-ID (ISR) of the National Synchrotron Light Source II, a U.S. Department of Energy (DOE) Office of Science User Facility operated for the DOE Office of Science by Brookhaven National Laboratory under Contract No. DE-SC0012704. Work supported by Brookhaven Science Associates, LLC
under Contract No. DE-SC0012704 with the U.S. Department of Energy. Work supported in part
by the U.S. Department of Energy under contract number DE-AC02-76SF00515 through FWP100903. This work was supported by the U.S. National
Science Foundation Grant No. PHY-1549132, the Center
for Bright Beams. This project has received funding from the European Union’s Horizon 2020 research and innovation programme under grant agreement No 101017902. 

\bibliography{references}




\end{document}